# Significant efficiency enhancement in thin film solar cells using laser beam-induced graphene transparent conductive electrodes

**L. V. Thekkekara,**[a*] **Bouyan Cai**[b]

[a] School of Science, RMIT University, Melbourne, Australia, 3000
[b] Nanophotonic Research Center, College of Optoelectronic Engineering, Shenzhen University, Shenzhen, China, 518060

**Abstract**. Thin film solar cells have been attractive for decades in advanced green technology platforms due to its possibilities to be integrated with buildings and on-chip applications. However, the bottleneck issues involved to consider the current solar cells as a major electricity source includes the lower efficiencies and cost-effectiveness. We numerically demonstrate the concept of the absorption enhancement in thin-film amorphous silicon solar cells using the laser beam-induced graphene material based on the insensitive polarization space-filling fractal design as transparent conductive electrodes. With the optimization of parameters such as thickness, width, and period of fractals, an enhancement of photocurrent generation of solar cells by a factor of 24.5% is achieved compared to reference solar cell with a traditional ITO.

## 1. INTRODUCTION

Flexible thin-film amorphous silicon (a-Si) solar cells are gaining much attention due to the progress of innovative portable and integrable technologies [1] which are in addition related to the inexpensive a-Si material [2]. The transparent conductive films (TCFs) with broadband transmission, excellent mechanical properties, polarization independence and light trapping properties play a significant part in the efficiency enhancement of the solar cells [3-4]. The commonly used TCFs in the current solar cells are based on transparent conductive oxides (TCOs), however the fragile nature and cost of the TCOs demand for new efficient, flexible designs and materials for extracting more light-generated electron-hole pairs in the solar devices [5].

Recent developments based on the optimized nanostructures such as the square grids and gratings as the transparent conductive electrodes (TCE) [6] can be a potential solution to overcome the issues, for example, replacement of expensive TCOs with cost-effective materials. In addition, tunable parameters like period, width, and thickness of nanostructures provide the efficient manipulation of photons and electrons in the solar cells by manipulating the transparency and the sheet resistance [7]. The light scattering and coupling offered by these nanostructures result in the enhancement of absorption in the active layer of the solar cells, which contributes to the solar energy to electricity conversion leading to the efficiency improvement [8-11]. The current nanostructure-based TCEs are suffering from the reduced transmission due to the polarization dependence which leads to the performance loss in the solar cells [12]. Recent reports on the application of fractal families like space filling curves [13] to achieve independent polarization TCEs will be a promising direction for transparent conductive electrodes.

Metals like gold, silver, and copper are generally used for the current TCE nanostructures, but the material cost of gold and silver are comparable to that of ITO [6]. In addition, the materials like copper suffer from the oxidation which makes it unrealistic for the practical applications [14]. The development of the cost-effective two-dimensional materials like reduced graphene oxides (rGOs) gives the provision for large area flexible TCE fabrication without the performance loss [15]. Among different reduction methods, photoreduction of GOs using direct laser writing method provides a single step fabrication of desired nanostructure pattern with the tunability of the resulted laser-induced graphene (LIG) properties like the improvement of electrical conductivity based on the laser irradiation parameters [16-17]. The major drawback of the LIG material to be considered as TCEs is the material light absorption loss depending on the reduction extent in the visible region [18].

Here we demonstrate the finite domain time domain simulation (FDTD) studies on the use of fractal designs for LIG TCEs to overcome the issues of polarization dependent transmission loss in the TCEs along with the compensation for the visible light absorption loss in the LIG material. In addition, the influence of the designed patterns on the light absorption enhancement of active layer in the thin film solar cells are considered. The optimization of the various fractal space-filling curves was reported in our previous literature [19].

## 2. DEVICE STRUCTURE AND SIMULATION

The details of the device structure are presented in Fig. 1. We consider three different LIG TCE nanostructure designs including linear gratings, square grids and fractals in comparison with ITO film on a thin-film a-Si solar cell using a commercial software package (Lumerical FDTD and Device Solutions). The thickness selected for the different layers in the a-Si thin-film solar cell such as ITO, a-Si, ZnO: Al and Ag film throughout the studies are as follows: 80 nm, 300 nm, 130 nm and 200 nm. For the optical simulation, a plane wave in the wavelength range from 300 to 1200 nm is used as the light source with polarization along the x-axis direction and propagation direction normal to the substrate plane.

Periodic boundary conditions are applied in the lateral direction for the simulation unit. Perfectly matched layer conditions are given in the perpendicular direction to keep off the noise issue. The mesh size of the simulation unit is set to be $\lambda/14$ in the material by taking $\lambda$ as smallest simulated wavelength. The refined mesh sizes of $\lambda/150$ are applied for the fractal designs and the interfaces until convergence is achieved. Two 2-D power monitors are placed above and below the a-Si layer to record the Poynting power transmitted through the network and to observe the remnant transmitted power through the a-Si layer.

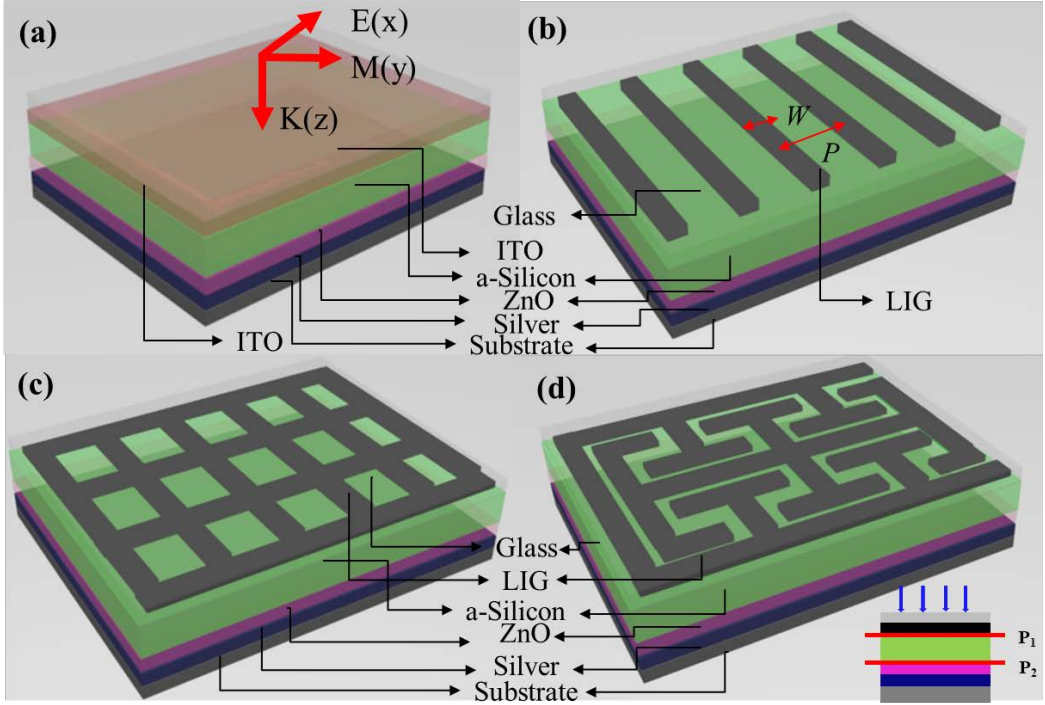

Fig. 1. The schematic setup of structures used for numerical simulations and light polarization directions. Thin-film a-Si solar cell with (a) ITO as TCO. (b) LIG grating as TCE. (c) LIG square grid as TCE. (d) LIG fractal as TCE. The width 'w' and period 'p' of the TCE is shown in (b), and the schematic of simulation unit with power monitors $P_1$ and $P_2$ used to obtain the transmission is shown in (d) as an inset.

The periodicities, *p* of the fractal designs used in the model varies from 150 to 600 nm, and the thickness of the LIG TCE is varied from 10 to 100 nm. The width, *w*, is 100 nm with an electrical conductivity of $10^3$ S/m from the experimental results during the femtosecond laser reduction of GOs under the repetition rate of 10 kHz with objective 100x and NA 1.4 using a wavelength of 800 nm. The refractive index of the LIG film is obtained from the ellipsometry measurement data using the analytical models as a function of wavelength (Fig. 2 (a)). The solar cell is encapsulated using the glass which serves as the antireflection coating [2]. The refractive indices of the different materials in a solar cell are obtained from the other literature [8].

The effect of the fractal structures on the a-Si solar cells is studied in detail by calculating the absorption of the a-Si, $A(\lambda)$, which is defined as the ratio of the power of the absorbed light, $P_{abs}(\lambda)$ to that of the incident light, $P_{in}(\lambda)$ within the a-Si film obtained from the transmission monitors in the simulation using the formula [22],

$$A(\lambda) = \frac{Pabs(\lambda)}{Pin(\lambda)} \quad (1)$$

The power absorbed can be written as

$$P_{abs}(\lambda) = 1/2\omega\varepsilon''|E|^2 \quad (2)$$

where $\omega$ is the frequency and $\varepsilon''$ is the imaginary part of the permittivity, and E is the electric field. The absorption enhancement, G in the a-Si layer is estimated with the fractal and grating TCE structures for different thickness and periods using the equation,

$$G = A_{TCE\ structures}(\lambda) / A_{ref}((\lambda) \quad (3)$$

where $A_{TCE}(\lambda)$ represents the absorption of the a-Si film with TCE fractal structures and $A_{ref}(\lambda)$ is the absorption in the a-Si film with TCE grating structures and so by integrating the absorption with the AM1.5 solar spectrum, we can get the area independent short circuit current density ($J_{SC}$), assuming all generated electron-hole pairs contribute to the photocurrent.

$$J_{sc} = e \int \left(\frac{\lambda}{hc}\right) A(\lambda) I_{AM1.5}(\lambda) d\lambda \quad (4)$$

where $e$ is the electron charge, $h$ is the Planck's constant, $c$ is the speed of light in free space, and $I_{AM1.5}$ is an AM1.5 solar spectrum.

The electron-hole pair (EHP) generation and recombination across the layers due to the wavelength-dependent absorption of incident light is essential to obtain an efficient quantum efficiency [23] of the solar cell. FDTD simulations cannot consider the factors mentioned above which necessitate a 3D model for the minority and majority carrier transport across the p and n junction based on Drift-diffusion (eqns. 5, 6) and Poisson's equation (eqn. 7) [24] with a finite mesh element discretization using a Dirichlet boundary and steady-state condition.

$$J_n = q\mu_n nE + qD_n \nabla_n \tag{5}$$

$$J_p = q\mu_p pE + qD_p \nabla_p \tag{6}$$

$$-\nabla.(\varepsilon \nabla V) = q\rho \tag{7}$$

where $J_{n,p}$ is the current density (A/cm$^2$), $q$ is the positive electron charge, $\mu_{n,p}$ is the mobility of charges, $E$ is the electric field, $D_{n,p}$ is the diffusivity, n and p are the densities of electrons and holes, $\varepsilon$ is the electric permittivity, $V$ is the electrostatic potential, $\rho$ is the net charge.

The model considers the surface and bulk recombination across the layers with the light source of incidence of power, 100 mW/cm$^2$. The efficiency, $\eta$ of the solar cells is calculated using the formula,

$$\eta = \frac{P_{max}}{I_{AM1.5}} \tag{8}$$

where $P_{max}$ is the maximum power density obtained (mW/cm$^2$) from the solar cells with TCEs and $I_{AM1.5}$ is the light source of incidence.

**Results and Discussion**

It is well known that the excitation of fundamental modes of TM or TE between the nanogratings or nanoholes leads to the enhanced light transmission [25]. The fractal design and arrangement consist of vertical and horizontal gratings which enable the stronger electromagnetic coupling and the energy transfer between both TE and TM modes leading to the polarization independent higher transmission [13]. In addition, the light trapping properties which are known for the fractal design contributed to the improvement [14].

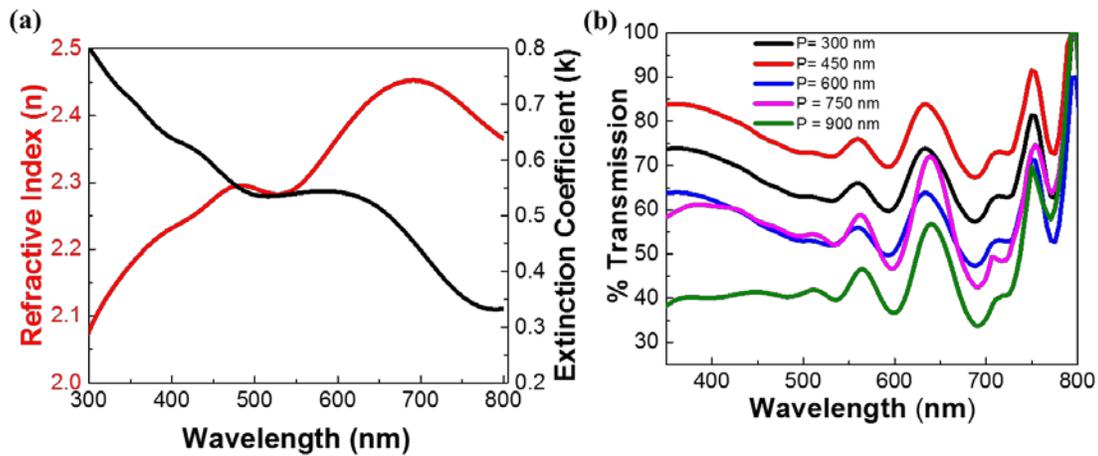

Fig. 2. (a) Refractive index and extinction co-efficient of LIG obtained from experimental results. (b) Transmission obtained after the LIG fractals TCEs by placing a 2D power monitor, $P_1$.

Fig. 2(b) represents the transmission obtained from a 2D power monitor after the LIG TCE. We can find that the transmission in the UV-visible region with the optimized LIG fractal TCE of thickness, 100 nm, and periodicity, 450 nm is around 80% while the transmissions with other LIG grating and grid parameters are decreased due to the TCE material absorption loss. The absorption peaks illustrated in the Fig. 2 (b) result from the LIG material [19]. In summary, the antireflection properties of the LIG material also contributed to the improvement in transmission [26]. Except for the optimization of the fractal TCEs, the different TCE designs are also presented in Table. 1 for a comparison of various TCE patterns under the TE, TM polarizations and averaged conditions.

| Type of TCEs | With antireflection coating $J_{sc}$ (mA cm$^{-2}$)  | | |
|---|---|---|---|
| | TE | TM | Average |
| ITO | | | 11.30 |
| LIG Grating | 13.99 | 14.18 | 14.08 |
| LIG Square grids | 13.98 | 14.12 | 14.05 |
| LIG Fractal | 14.20 | 14.20 | 14.20 |
| Ag nanowires [8] | | | 14.15 |

Table 1. The *Jsc* generated with the various types of TCE electrodes using the FDTD optical simulation under different polarization conditions.

It can be concluded that the absorption in the a-Si layer is higher with LIG fractal TCE and the absorption enhancement is broadband, ranging from UV to near infrared as shown in Fig. 3(a) whereas the other solar cells with TCEs made of polarization averaged gratings and square grids have lower absorption. Absorption enhancement is calculated using the eqn. 3 concerning ITO based a-Si solar cells and the results shows (Fig. 3 (b)) an improvement up to 1.3% for the LIG fractal TCEs in the UV-visible region.

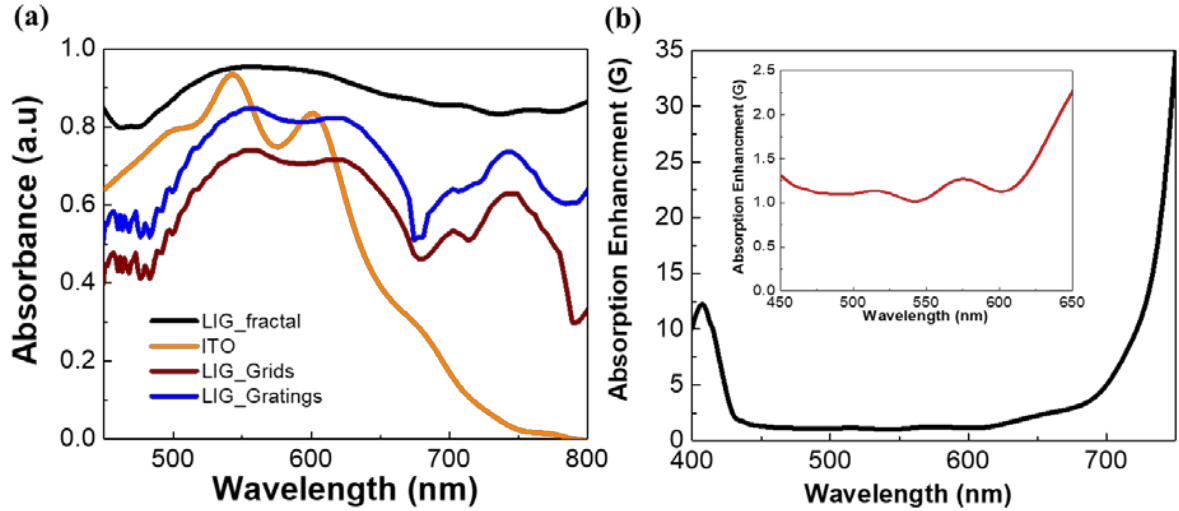

Fig. 3. (a) Absorption in the thin film a-Si solar cell. The periodicities considered for the TCE designs are 450 nm, the width is 100 nm and thickness is 80 nm (b) Calculated absorption enhancement, G in a-Si layer using LIG fractal TCE of period 450 nm, width 100 nm, and thickness 80 nm.

To further demonstrate the performance improvement of the solar cell device with our designed TCE nanostructures, we calculated the $J_{sc}$ value with different TCE patterns as shown in Fig. 4. A notable improvement of around 24.5% $J_{sc}$ enhancement can be achieved with the optimized LIG fractal TCE structure with a periodicity of 450 nm and a thickness of 80 nm. The significant Jsc enhancement can be ascribed to the polarization insensitiveness along with the light trapping properties due to the specific arrangement of the fractal design. In addition, the antireflection properties of LIG material arising from the refractive index matched between the a-Si and the air can also lead to the enhancement in the short wavelength range.

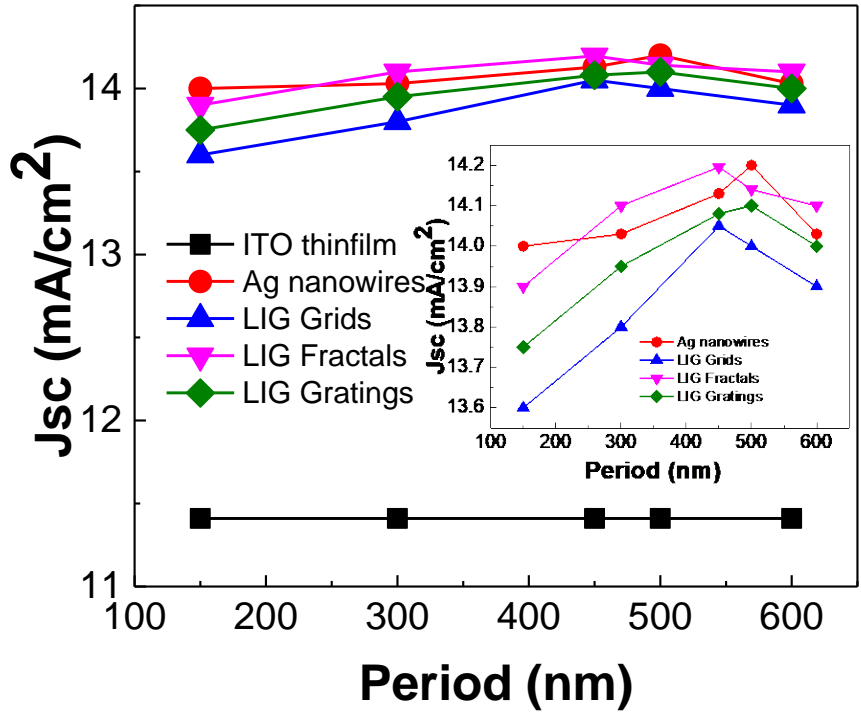

Fig. 4. $J_{sc}$ value improvement in the thin-film **a-Si** solar cell with LIG TCEs of thickness 80 nm, period 450 nm and width 100 nm in comparison with ITO film. The Jsc values with Ag grids are taken from the reference[8].

A comparison between the optical and electrical properties of the solar cell are presented in Table 2 which shows the effect of recombination in the efficiency of solar electricity generation in comparison to the optical absorption consideration along in the numerical studies.

|  | With antireflection coating | | | |
|---|---|---|---|---|
|  | Jsc (mA cm$^{-2}$) | | Efficiency, η | |
|  | Optical | Electrical | Optical | Electrical |
| Bare a-Si solar cell with ITO | 11.41 | 10.1 | 10.0 | 9.5 |
| Grating | 14.08 | 11.5 | 11.5 | 10.9 |
| Square grids | 14.05 | 11.1 | 11.2 | 10.5 |
| Fractal | 14.20 | 12.2 | 12.0 | 11.4 |
| Ag nanowires [8] | 14.14 | - | | |

Table 2. Comparison between the properties of the thin-film a-Si solar cell using optical and electrical analytical modeling.

## 3. CONCLUSION

In summary, this paper introduces a novel approach to enhance the broadband absorption in the a-Si thin-film with the application of fractal TCE design using the LIG material. The proposed design results in the *Jsc* enhancement of 24.5% in the a-Si thin film solar cells comparing to the traditional ITO TCOs. In addition, the simulation results pave a direction to overcome the absorption issues in the LIG material which opens the prospectus of cost-effective material for the transparent optoelectronic devices.

**Acknowledgment.** L. V. T acknowledges RMIT University for the RPIS scholarship and center for Micro-Photonics, Swinburne University for the computational facilities. B. C. thanks the support from the National Natural Science Foundation of China (NSFC) (61427819, 61605064), Leading talents of Guangdong province program (00201505). Science and Technology Innovation Commission of Shenzhen (KQCS2015032416183980), Natural Science Foundation of SZU (000011, 2017047), Natural Science Foundation of Guangdong Province, China (2016A030312010, 2016A03031008).